\documentclass[a4paper]{spie}

\usepackage{amsfonts,amsmath,graphicx,algorithmic,caption,subfig,algorithm,algorithmic}

\begin{document}

\title{Compressive imaging using fast transform coding}
\author[a]{Andrew Thompson}
\author[b]{Robert Calderbank}
\affil[a]{Mathematical Insititute, University of Oxford, UK}
\affil[b]{Department of Electrical and Computer Engineering, Duke University, USA}

\authorinfo{Further author information: (Send correspondence to A.T.)\\A.T.: E-mail: thompson@maths.ox.ac.uk\\  R.C.: E-mail: robert.calderbank@duke.edu}

\newcommand{\mK}{\mathcal{K}}
\newcommand{\RR}{\mathbb{R}}
\newcommand{\ZZ}{\mathbb{Z}}
\newcommand{\argmin}{\operatornamewithlimits{argmin}}
\newtheorem{thm}{Theorem}

\maketitle

\begin{abstract}
We propose deterministic sampling strategies for compressive imaging based on Delsarte-Goethals frames. We show that these sampling strategies result in multi-scale measurements which can be related to the 2D Haar wavelet transform. We demonstrate the effectiveness of our proposed strategies through numerical experiments.
\end{abstract}


\section{Introduction}

Recent advances in sparse inverse problems, often referred to as compressive sensing~\cite{donoho,robust}, have led to new methods and technologies for compressive image reconstruction. The central idea is to design encoded, subsampled linear measurements which allow for accurate optimization-based decoding which assumes sparsity in some transform domain such as wavelets or curvelets~\cite{mallat}. Applications of interest include fluorescence microscopy~\cite{studer}, infrared microscopy~\cite{gallet} and astronomy, for example the single-pixel telescopic system~\cite{yu}.

Intensity measurements, in which a pixel array is sampled using a binary `on-off' mask, are often considered an attractive choice. For example, in the case of the single-pixel camera~\cite{spc}, a spatial light modulator (SLM) in the form of a digital micromirror device is typically used for image acquisition, and binary measurements are simplest to encode into hardware. Meanwhile, it is important for the computational efficiency of the optimization-based reconstruction that the measurement matrix can be implemented as a fast transform. The Walsh-Hadamard Transform (WHT)~\cite{walsh_paley}, which involves a measurement matrix consisting of $\pm 1$ entries, is therefore an attractive solution, and so measurement schemes are often based upon it (see for example each of the references given above).

One of the central tenets of compressive sensing is that uniform, random, subsampling is optimal for reconstructing signals whose sparsity pattern is unstructured~\cite{robust}. However, wavelet transforms have an inherent multi-scale structure, and the sparsity patterns associated with the wavelet transforms of natural images are in consequence highly structured, with signal energy being concentrated in the coarse (low frequency) scales. In fact, the WHT itself has a multi-scale interpretation, and it has now been established that improvements in reconstruction accuracy result from subsampling the WHT at a variable rate; and in particular, more aggressively at finer (higher frequency) scales~\cite{quest}. 

Practical schemes incorporating the above principles typically involve randomly subsampling the WHT coefficients, while varying the subsampling rate with scale~\cite{quest}. Implementation of this approach requires the locations of the selected random coefficients to be stored and the full WHT to be repeatedly computed within the reconstruction algorithm. 

In this paper, we consider an entirely deterministic approach to variable, multi-scale, $\pm 1$ sampling. The scheme leads to comparable reconstruction performance, but without the need to rely upon randomness. As a result, no locations of randomly selected coefficients need to be stored, and full WHTs can be replaced with more efficient transforms which directly compute the subsampled coefficients. The sampling scheme is based upon a family of matrices that have been shown both theoretically and empirically to be effective measurement matrices for compressive sampling: Delsarte-Goethals frames~\cite{strip,structured}. We will consider one particular instantiation of Delsarte-Goethals frames: {\em real Kerdock matrices}. The contributions of this paper can be summarized as follows.

\begin{itemize}
\item We show that applying a Kerdock matrix to a signal results in multi-scale measurements, where the measurements at a given scale sample only a single scale of the {\em Haar wavelet transform} of the signal. In other words, the Kerdock transform is scale-preserving.
\item Based on these new insights, we propose a new deterministic strategy for variable, multi-scale, $\pm 1$ sampling. We present experimental evidence that the new strategy leads to improved image reconstruction performance compared to the uniformly subsampled WHT.
\end{itemize}

The structure of the rest of the paper is as follows. In Section~\ref{kerdock}, we give details on the real Kerdock transform, and in Section~\ref{scale}, we establish the scale-preserving properties of the transform and make the connection to the Haar wavelet transform. We then propose our new deterministic multi-scale sampling strategy in Section~\ref{strategy}. Numerical image reconstruction experiments can be found in Section~\ref{experiments}, comparing the deterministic approach with other subsampling strategies. 

\section{The Kerdock transform}\label{kerdock}

In this section, we give the essential form of the Kerdock transform and refer the interested reader to~\citenum{strip,structured} for further technical details. We define a $2^m\times 2^{m+p}$ Kerdock matrix $K^{m,p}$ as
$$K^{m,p}=\begin{bmatrix}D^1 H_m&D^2 H_m&\ldots&D^{2^p}H_m\end{bmatrix},$$
where $H_m$ is a $2^m\times 2^m$ normalized Hadamard matrix and $D^i$, for $i\in\{1,2,\ldots,2^p\}$, is a diagonal matrix with $\pm 1$ entries on the diagonal. These diagonal entries are determined by binary quadratic codes known as Kerdock codes (see~\citenum{strip,structured} for further details).

The Kerdock matrix $K^{m,p}$ is a union of orthobases, and it is instructive to consider how it operates on a vector divided into $2^p$ subvectors of length $2^m$. Writing 
$$x=\begin{bmatrix}(x^1)^T&(x^2)^T&\ldots&(x^{2^p})^T\end{bmatrix}^T,$$
we have 
$$K^{m,p}x=\begin{bmatrix}D^1 H_m&D^2 H_m&\ldots&D^{2^p}H_m\end{bmatrix}\begin{bmatrix}x^1\\x^2\\ \vdots\\x^{2^p}\end{bmatrix}=\displaystyle\sum_{i=1}^{2^p}D^i H_m x^i.$$
In other words, it is a weighted sum of the WHTs of the subvectors $\{x^i\}$. It follows immediately that there exists an $\mathcal{O}(m\cdot 2^{m+p})$ transform for computing a matrix-vector product with $K^{m,p}$: divide the input vector into subvectors, apply the WHT to each, then apply the necessary componentwise sign changes to the resulting outputs and sum them. We will call this transform the 1D Kerdock transform, and denote it by $\mK^{m,p}$.

The parameter $p$ controls the compression factor of the transform, which is $2^p$. The ability to vary the parameter $p$ for different wavelet scales will be essential for designing a variable, multi-scale sampling strategy. We note that there is an upper bound on $p$ imposed by the underlying Kerdock codes: we must have $p\le m-1$ (see~\citenum{strip,structured}). 

The extension to a 2D Kerdock transform is achieved using the usual Cartesian product. Given a $2^{m+p}\times 2^{m+p}$ image $X$, define the 2D Kerdock transform $\mK_{(2)}^{m,p}$ to be
$$\mK_{(2)}^{m,p}(X) = K^{m,p}X(K^{m,p})^T.$$

\section{The scale-preserving property of the Kerdock transform}\label{scale}

The aim of this section is to show that applying a Kerdock matrix to a signal results in multi-scale measurements, where the measurements at a given scale sample only a single scale of the Haar wavelet transform of the signal. In other words, the Kerdock transform is scale-preserving. We restrict our focus in this section to the 1D Kerdock transform, though the analysis extends naturally to the 2D Kerdock transform and images.

Our starting point is a striking result about the relationship between the Walsh-Hadamard Transfrom (WHT) and the Haar wavelet transform. Given $m\geq 0$, let $H_m$ be the $2^m\times 2^m$ Hadamard matrix with columns in dyadic/Paley order~\cite{walsh_relations,walsh_paley}, and let $\Psi_m$ be the matrix whose rows are the Haar wavelet basis elements ordered in the usual way from coarse to fine scales, so that multiplication by $\Psi_m$ gives the discrete Haar wavelet transform. The matrix product $H_m\Psi_m^T$ has the following block-diagonal decomposition.  

\noindent\textbf{Theorem 1.}
\begin{equation}\label{coherence}
H_m\Psi_m^T=\begin{bmatrix}
1&&&&\\
&H_0&&&\\
&&H_1&&\\
&&&\ddots&\\
&&&&H_{m-1}\end{bmatrix}.
\end{equation}

The result appears implicitly in the literature, for example in~\citenum{walsh_relations,fino}, but appears explicitly here for the first time, to the authors' best knowledge. A proof is given in Appendix~\ref{proof}. The result can also be viewed as a statement about the mutual coherence of the Hadamard and Haar bases. Since the magnitude of the coefficients decays with scale, it is an example of  
the asymptotic incoherence described in~\citenum{quest}.

Now consider the Kerdock transform of a signal $x\in\RR^{2^{m+p}}$ (see Section~\ref{kerdock}), and write 
$$x=\begin{bmatrix}(x^1)^T&(x^2)^T&\ldots&(x^{2^p})^T\end{bmatrix}^T$$
for the decomposition of $x$ into $2^p$ subsignals, so that
$$y=\begin{bmatrix}D^1 H_m&D^2 H_m&\ldots&D^{2^p}H_m\end{bmatrix}\begin{bmatrix}x^1\\x^2\\ \vdots\\x^{2^p}\end{bmatrix}.$$
Let $w=\Psi_{m+p}x$ be the Haar wavelet transform of $x$, and write 
$$w=\begin{bmatrix}\nu&w_0^T&w_1^T&\ldots&w_{m+p-1}^T\end{bmatrix}^T$$
for its Haar wavelet decomposition by scales. Write $w^i=\Psi_m x^i$ for the Haar wavelet transform of each subsignal $x^i$, $i\in\{1,2,\ldots,2^p\}$, and 
$$w^i=\begin{bmatrix}\nu^i&(w_0^i)^T&(w_1^i)^T&\ldots&(w_{m-1}^i)^T\end{bmatrix}^T$$
for their respective Haar wavelet decompositions by scales. Also write
$$D^i=\begin{bmatrix}
1&&&&\\
&D_0^i&&&\\
&&D_1^i&&\\
&&&\ddots&\\
&&&&D_{m-1}^i\end{bmatrix}$$
for each diagonal matrix $D^i$, $i\in\{1,2,\ldots,2^p\}$, where the submatrices correspond in size to the wavelet scales. Then we have
$$\begin{array}{rcl}\mK^{m,p}(x)&=&K^{m,p}x=\begin{bmatrix}D^1 H_m&D^2 H_m&\ldots&D^{2^p}H_m\end{bmatrix}\begin{bmatrix}x^1\\x^2\\ \vdots\\x^{2^p}\end{bmatrix}\\
&=&\displaystyle\sum_{i=1}^{2^p}D^i H_m x^i\\
&=&\displaystyle\sum_{i=1}^{2^p}\begin{bmatrix}
1&&&&\\
&D_0^i&&&\\
&&D_1^i&&\\
&&&\ddots&\\
&&&&D_{m-1}^i\end{bmatrix}
\begin{bmatrix}
1&&&&\\
&H_0&&&\\
&&H_1&&\\
&&&\ddots&\\
&&&&H_{m-1}\end{bmatrix}
\begin{bmatrix}
\nu^i\\
w_0^i\\
w_1^i\\
\vdots\\
w_{m-1}^i
\end{bmatrix}\\
&=&\begin{bmatrix}\sum_{i=1}^{2^p}\nu^i\\
\sum_{i=1}^{2^p}D_0^i H_0 w_0^i\\
\sum_{i=1}^{2^p}D_1^i H_1 w_1^i\\
\vdots\\
\sum_{i=1}^{2^p}D_{m-1}^i H_{m-1} w_{m-1}^i
\end{bmatrix}.
\end{array}$$

But Haar wavelets have a nested property: if we divide a signal $x$ into $2^p$ subsignals $x^i$, the Haar wavelet coefficients of $x$ at scale $j+p$ are nothing other than the Haar wavelet coefficients of each of the subsignals at scale $j$, that is,
$$w_{j+p}=\begin{bmatrix}(w_j^1)^T&(w_j^2)^T&\ldots&(w_j^{2^p})^T\end{bmatrix}^T.$$
It follows that, writing
$$y=\begin{bmatrix}\mu&y_0^T&y_1^T&\ldots&y_{m-1}^T\end{bmatrix}^T$$
for the decomposition of the output measurements by scale, we have, for $j\in\{0,1,\ldots,m-1\}$,
$$y_j=\sum_{i=1}^{2^p}D_j^i H_j w_j^i=\begin{bmatrix}D_j^1 H_j&D_j^2 H_j&\ldots&D_j^{2^p} H_j\end{bmatrix}\begin{bmatrix}w_j^1\\w_j^2\\ \vdots\\w_j^{2^p}\end{bmatrix}=\begin{bmatrix}D_j^1 H_j&D_j^2 H_j&\ldots&D_j^{2^p} H_j\end{bmatrix}w_{p+j}.$$
The output measurements at a given scale $j$ are thus seen to be nothing other than a weighted sum of the Haar wavelet coefficients at scale $j+p$\footnote{It is apparent that each wavelet scale is sampled by a measurement matrix which itself looks very much like a Kerdock matrix, though strictly subcodes of a Kerdock code are not themselves Kerdock codes. However, in our numerical experiments in Section~\ref{experiments}, we replaced these subcodes with actual Kerdock codes for each scale, which is a departure from~\citenum{strip,structured}.}. 

\section{A deterministic strategy for multi-scale sampling}\label{strategy}

We can summarize the implications of the previous section as follows.

\begin{enumerate}
\item The Kerdock transform $\mK^{m,p}$ performs uniform sampling of the finest $m$ Haar wavelet scales and gathers no information from the first $p$ wavelet scales. It can therefore be combined with direct sampling of the coarsest $p$ wavelet scales to give a two-level sampling scheme. Note the desirable property that the coarsest $p$ scales, for which full samples are computed, are not further sampled unnecessarily by the Kerdock transform.
\item The subsampling factor $2^p$ can be chosen for each scale independently by computing measurements at each scale using whichever $\mK^{m,p}$ transform is desirable. An important remaining question is how to efficiently implement this multi-scale sampling scheme, and we leave this question as future work.
\end{enumerate}

Based on these observations, we propose a deterministic multi-scale sampling scheme for 1D signals.

\begin{algorithm}[h]\caption{Deterministic multi-scale sampling scheme.}
\label{our_strategy}
\textbf{Inputs:} Signal $x\in\RR^{2^m}$; sampling strategy $P\in\ZZ_+^m$.
\begin{algorithmic}
\STATE $y_1=\left\{\Psi x\right\}_1$
\FOR {$j=1:m$}
\FOR {$i=2^{j-1}+1:2^j$}
\IF {$P_j=0$} 
\STATE $y_i=\left\{\Psi x\right\}_i$
\ELSE 
\STATE $y_i=\left\{\mK^{m-P_j,P_j}(x)\right\}_i$
\ENDIF
\ENDFOR
\ENDFOR
\end{algorithmic}
\textbf{Outputs:} Measurements $y$.
\end{algorithm}

Note that the required user tuning is straightforward and intuitive: simply provide a vector $P$ of integers, the entries of which give the power of $2$ by which the signal will be subsampled at each scale. Let us write $\mK^P$ for this 1D multi-scale Kerdock transform, and $K^P$ for the corresponding measurement matrix. 

The extension to 2D images immediately follows as in Section~\ref{kerdock}. Given a $2^m\times 2^m$ image $X$, define the 2D multi-scale Kerdock transform $\mK_{(2)}^P$ to be
$$\mK_{(2)}^{P}(X) = K^{P}X(K^{P})^T.$$

\section{Numerical experiments}\label{experiments}

We present a comparison of image reconstruction using different schemes for $\pm 1$ sampling. We sample the 1024x1024 `Man' test image by taking $m=320^2$ samples (a subsampling factor of $10.24$). Writing $w\in\RR^n$ ($n=1024^2$) for the vectorized 2D Haar wavelet coefficients of the image, we can represent the linear measurements in the form $b=Aw\in\RR^m$, where $A$ is an $m\times n$ matrix. We then use the \texttt{spgl1} code~\cite{spgl1} to solve the optimization problem
$$\hat{w}:=\argmin_{w}\|w\|_1\;\;\;\textrm{subject to}\;\;b=Aw.$$
We measure accuracy of reconstruction using the signal-to-noise ratio (SNR), defined as
$$\textrm{SNR}:=10\log_{10}\frac{\|w\|_2}{\|\hat{w}-w\|_2}.$$

\begin{figure}[t!]
\centering
\subfloat[Original]{\includegraphics[width=0.4\textwidth]{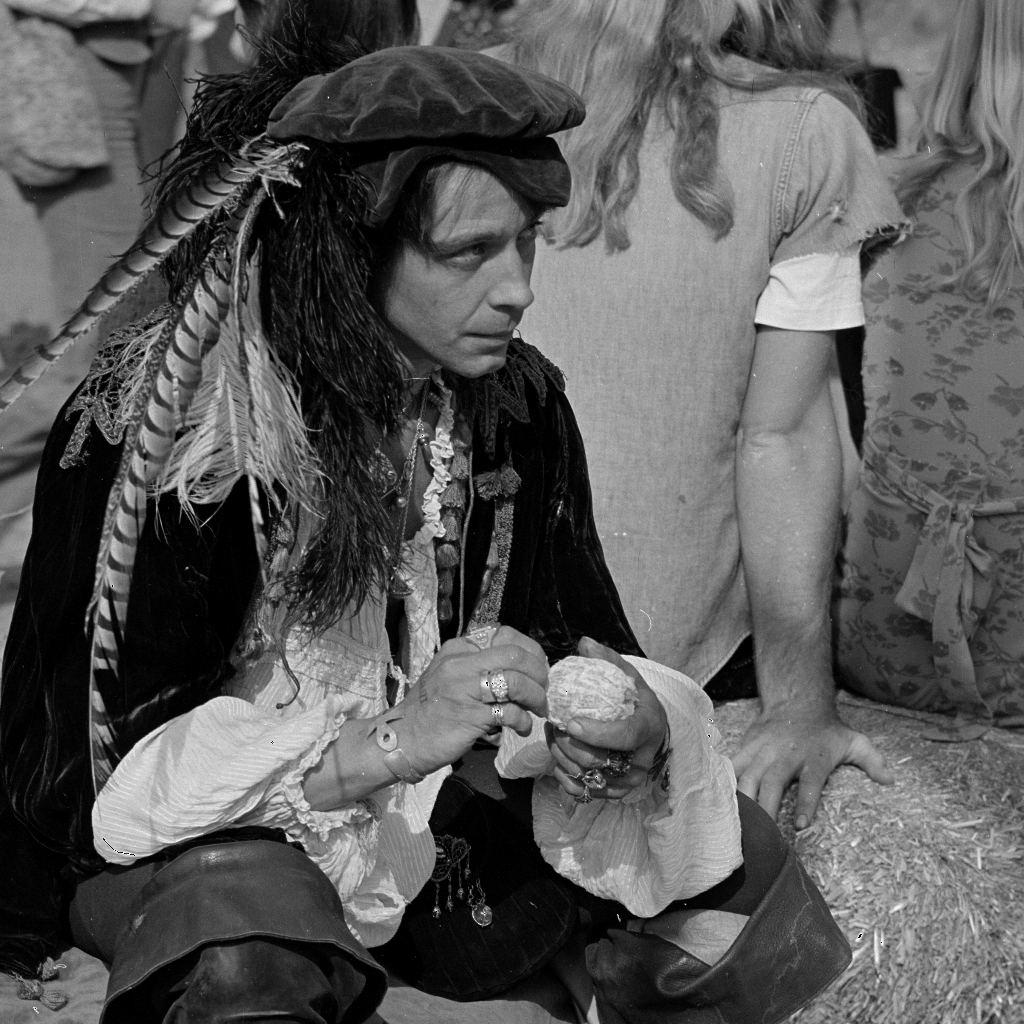}}\hspace{1cm}
\subfloat[Scheme 1 (SNR 2.7151)]{\includegraphics[width=0.4\textwidth]{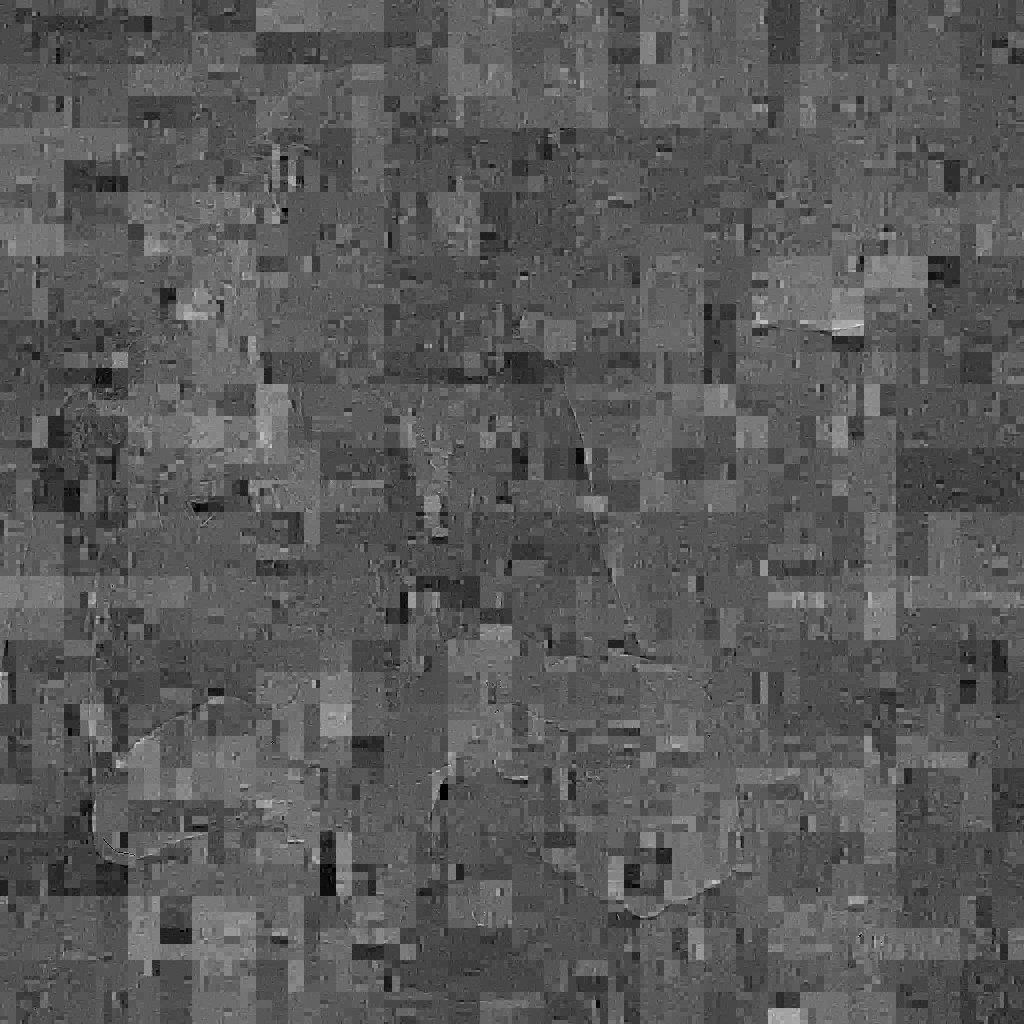}}\\
\subfloat[Scheme 2 (SNR 7.8643)]{\includegraphics[width=0.4\textwidth]{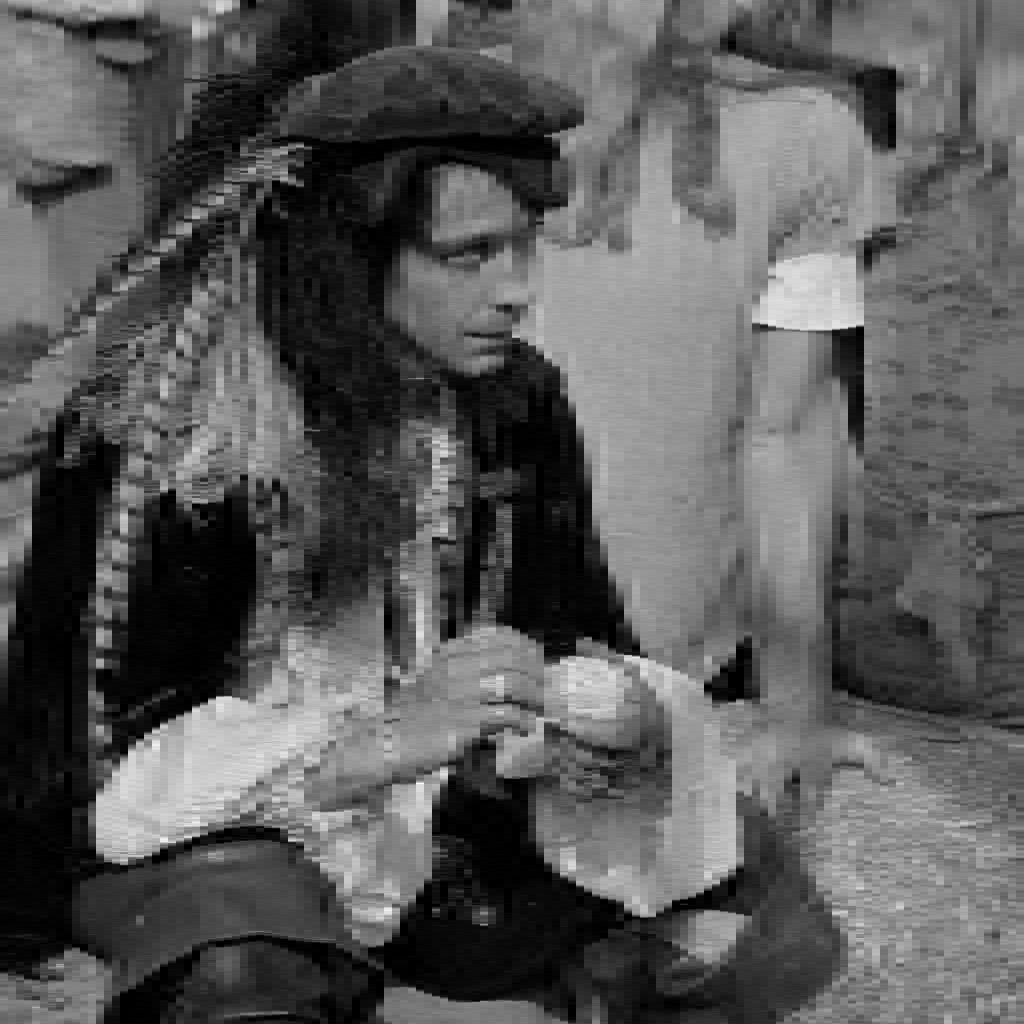}}\hspace{1cm}
\subfloat[Scheme 3 (SNR 8.6924)]{\includegraphics[width=0.4\textwidth]{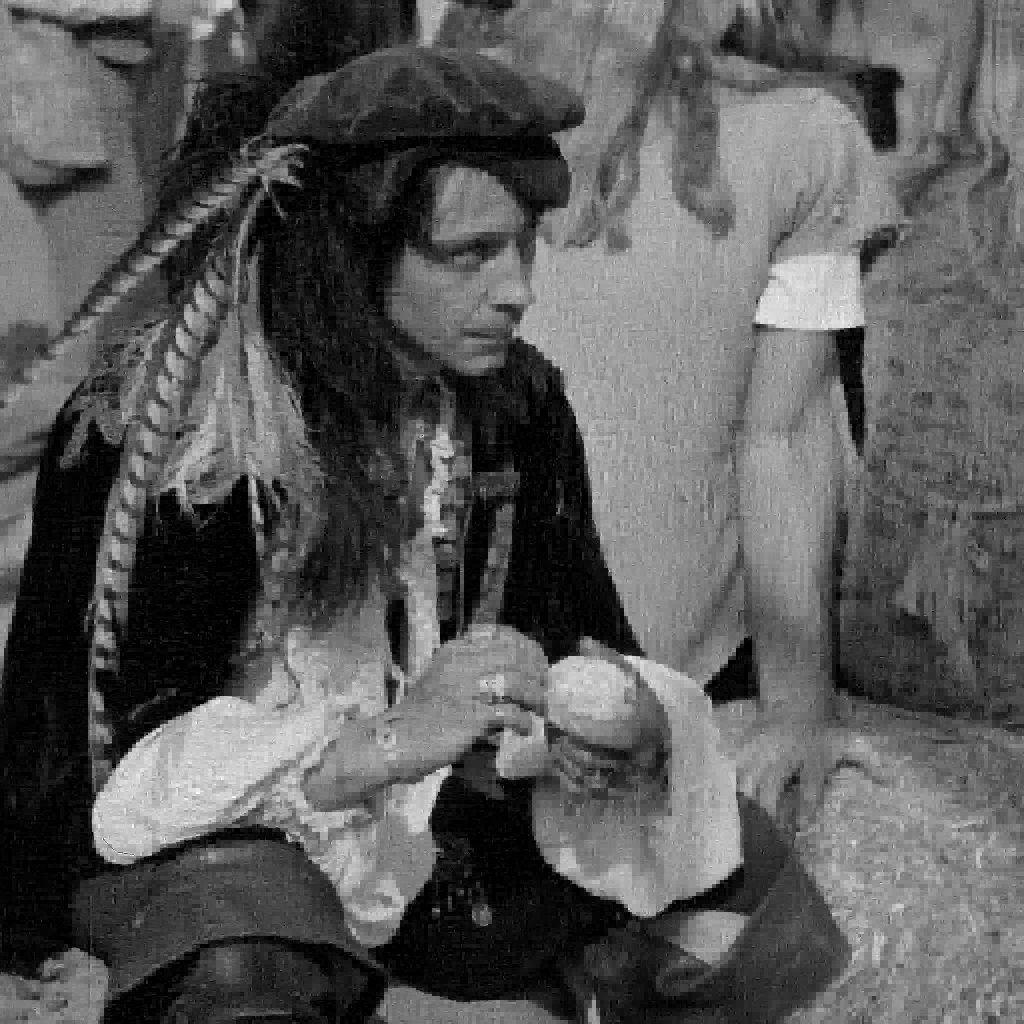}}\\
\caption{Images reconstructed by $l_1$-minimization from various $\pm 1$ sampling schemes.}
\label{image_plots}
\end{figure}

We present results for three sampling schemes in Figure~\ref{image_plots}. Schemes 1 and 2 are subsamplings of the 2D WHT, taking $m=320^2$ randomly chosen coefficients (scheme 1) and the first (lowest frequency) $m=320^2$ coefficients (scheme 2) respectively. We see that the performance of random subsampling (scheme 1) is, as the research consensus would now expect, disastrous, while the low frequency sampling (scheme 2) represents a baseline with which to compare. 

Scheme 3 is a version of the Kerdock sampling scheme described in this paper, in which we directly sample the first $7$ scales of the WHT, and then use Kerdock matrices to subsample scales $8$ to $10$, using $p=1$ for scale $8$, $p=2$ for scale $9$, and $p=3$ for scale $10$ (see Section~\ref{kerdock}), again giving $m=320^2$. We observe that the reconstruction accuracy using scheme 3 represents a significant improvement over scheme 2, demonstrating the effectiveness of a deterministic multi-scale sampling strategy with a number of different subsampling factors across scales (in this case four).

\appendix

\section{Proof of Theorem 1}\label{proof}

Given $m\geq 0$, the $2^m\times 2^m$ Hadamard matrix with columns in dyadic/Paley order~\cite{walsh_relations,walsh_paley}, $H_m$, is defined by the recursion
\begin{equation}\label{walsh_dyadic}
H_0:=1;\;\;\;\;H_{m+1}=\frac{1}{\sqrt{2}}\left[\begin{array}{c}H_m\otimes\left(\begin{array}{ll}1&1\end{array}\right)\\H_m\otimes\left(\begin{array}{ll}1&-1\end{array}\right)\end{array}\right]\;\mbox{for}\;m\geq 0,
\end{equation}
where $\otimes$ denotes the usual Kronecker product. Given $m\geq 0$, the $2^m\times 2^m$ Haar transform matrix~\cite{walsh_paley}, $\Psi_m$, may be defined by the recursion
\begin{equation}\label{haar}
\Psi_0:=1;\;\;\;\;\Psi_{m+1}=\frac{1}{\sqrt{2}}\left[\begin{array}{c}\Psi_m\otimes\left(\begin{array}{ll}1&1\end{array}\right)\\I_m\otimes\left(\begin{array}{ll}1&-1\end{array}\right)\end{array}\right]\;\mbox{for}\;m\geq 0,
\end{equation}
where $I_m$ is the $2^m\times 2^m$ identity matrix.

The proof of Theorem 1 is by induction on $m$. Noting that (\ref{coherence}) holds trivially for $m=0$, assume (\ref{coherence}) holds for $m=r\geq 0$. Using (\ref{walsh_dyadic}) and (\ref{haar}), and by symmetry of $\Phi_m$, we have
$$\Phi_{r+1}\Psi_{r+1}^T=\frac{1}{2}\left[\begin{array}{c}\Phi_r\otimes\left(\begin{array}{cc}1&1\end{array}\right)\\\Phi_r\otimes\left(\begin{array}{cc}1&-1\end{array}\right)\end{array}\right]\left[\begin{array}{ll}\Psi_r^T\otimes\left(\begin{array}{c}1\\1\end{array}\right)&I_r\otimes\left(\begin{array}{c}1\\-1\end{array}\right)\end{array}\right]=\left[\begin{array}{cc}\Phi_r\Psi_r^T&0\\0&\Phi_r\end{array}\right],$$
and (\ref{coherence}) now follows for $m=r+1$, and hence for all $m\geq 0$ by induction.\hfill$\Box$

\end{document}